# The possibility of noninvasive micron high energy electron beam size measurement using diffraction radiation.

G. Naumenko
*Nuclear Physics Institute at Tomsk Polytechnic University, Tomsk, Russia.*
E-mail: naumenko@npi.tpu.ru

**Abstract**
Treatments of the usage of optical diffraction radiation from the relativistic electrons moving though a conductive slit for the noninvasive transverse beam size measurement (for example [1]) encounter hard limitation of the method sensitivity for the electron energy larger than 1 GeV. We consider in this article a possibility of application in a diffraction radiation technique the artificial phase shift, which can take place when transverse electron position varies. This allows us to realize the nonivasive measurements of transverse size of supper-relativistic electron beams with the small emittance.

**Introduction**

Recently the projects on the creation of the noninvasive diagnostics methods using optical diffraction radiation (ODR) of electrons, passing through conductive slit (Fig.1), based on the work [2], are in progress (for example [1]). According this work the real criteria for the beam size assess is the relation $Y_\sigma/Y_{max}$ between minimum and maximum of ODR intensity in the angular distribution of the parallel polarization radiation component (see Fig.2) .

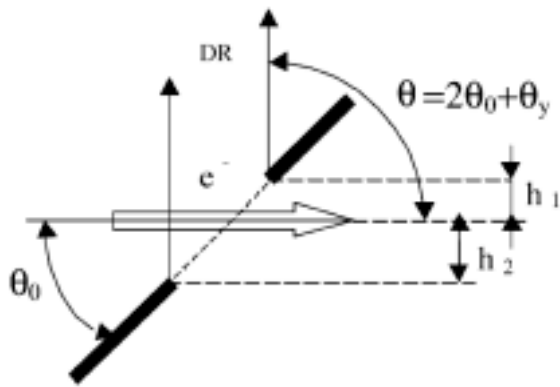

Fig.1. Geometry of ODR from slit from [1].

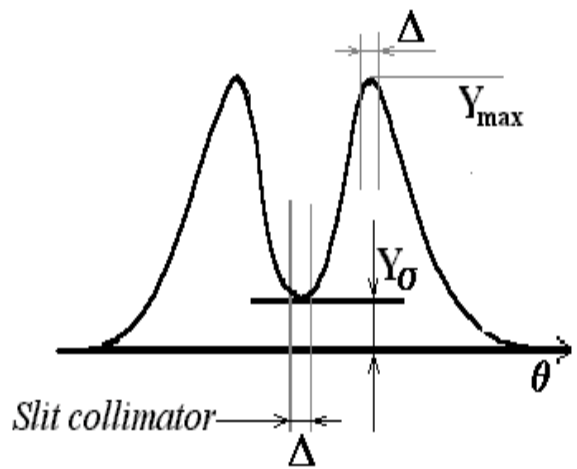

Fig.2. ODR intensity angular distribution of the polarization radiation component in reflection plane. $Y_\sigma$ is the response on the beam size.

Optimal condition for beam size measurement is the usage of narrow slit collimator the same for both $Y_\sigma$ and $Y_{max}$, as it is shown in Fig.2. In this relation in a first approach only $Y_\sigma$ is sensitive to a beam size. As it is small, we must be sure, that $Y_\sigma$ is sizable in comparison to the background.

Any electron beam is not free from bending magnets, steering magnets and magnetic lenses. Synchrotron radiation from edges of these elements will reflect from ODR slit target and will propagate to a detector as well, as the ODR. This radiation we will denote the synchrotron radiation background (SRB). It was shown in [4], that SRB differential intensity $Y_{SRB}$ may reach the value of differential intensity $Y_{TR}$ of transition radiation (TR) from conductive target, and angular properties of SRB are similar to ODR one. So it is impossible to separate ODR and SRB. On the other hand ODR intensity in a maximum of angular distribution is less than OTR intensity. For example maximum ODR intensity from the slit target with slit width $0.2\gamma\lambda$ comprises 0.6 of OTR intensity. Here $\gamma$ is Lorenz-factor of electron, $\lambda$ is a radiation wavelength. The common property of these radiation types is the dependence of the maximum radiation intensity on the Lorenz-factor for similar conditions as $\gamma^2$. Also it should be noted that SRB depends on the beam line tuning and is not repetitive.

It is convenient to consider the relation of ODR and SRB intensity to the TR intensity for the same conditions because for a small width of slit collimator this relation does not depend on the collimator width with a good accuracy. We will consider the radiation polarization component in reflection plane in a narrow slit collimator, which is perpendicular to the reflection plane (see Fig.1). For TR measurement the position of slit collimator in the maximum of TR angular distribution assumed. We suppose nice conditions, when $Y_{SRB}/Y_{TR} \approx 0.05$. So we require the $Y\sigma$ the same order as $Y_{SBR}$ or lager. For calculation we used the expression for ODR differential intensity from slit target from [1]. There was performed integral over an angular distribution in the slit collimator and integral over the Gaussian transversal distribution of electrons in the beam with parameter $\sigma$. The relation to the TR yield for the same condition we can obtain in approach of a narrow slit collimator a simple formula:

$$\frac{Y_\sigma}{Y_{OTR}} \approx 4\pi^2 \left(\frac{\sigma}{\gamma\lambda}\right)^2 e^{-2\pi\left(\frac{a}{\gamma\lambda}\right)},$$

where $a$ is a slit width in slit target. Maximum relation we obtain if we minimize a value of $a$.

$$\frac{Y_\sigma}{Y_{OTR}} \approx 4\pi^2 \left(\frac{\sigma}{\gamma\lambda}\right)^2 \qquad (1)$$

Let us underline, that we consider most favorable conditions for beam size measurement using ODR.

Resolving equation (1) in respect to the beam size, we obtain a limitation of transversal beam size for measurement using ORD, caused by SRB:

$$\sigma \approx \frac{\gamma\lambda}{2\pi} \sqrt{\frac{Y_\sigma}{Y_{OTR}}} \qquad (2)$$

For example:
1. Conditions of KEK ATF (Japan): $\gamma=2500$, $\lambda \approx 0.5$mcm, $Y_\sigma/Y_{TR} \approx 0.05$ → $\sigma \approx 40$mcm.
2. Condition of SLAC FFTB: $\gamma=60000$, $\lambda \approx 0.5$mcm, $Y_\sigma/Y_{TR} \approx 0.05$ → $\sigma \approx 1$mm.

A beam size response for SLAC conditions with beam size=10mcm may be obtained from (1):

$$\frac{Y_\sigma}{Y_{OTR}} \approx 4\pi^2 \left(\frac{10\,mcm}{60000 \cdot 0.5\,mcm}\right)^2 \approx 5 \cdot 10^{-6}$$

It is clean, that not any experiment can provide such accuracy.

We come to recognize that the method suggested in [2] don't allows us to obtain a necessary sensitivity for measurement of fine focused high energy beam transversal size.

## Beam size measurement using diffraction radiation phase shift.

. We suggest to apply the artificial phase shift which can take place when transverse electron position varies. This method ensures much more sensitivity and resolution, which is limited only by the radiation wavelength. For the realizing of the phase shift, depending on the transverse electron position, we suggest to turn at the small angle $\alpha$ (see Fig.3) around the vertical axis both semi-planes forming the slit target, relative one to other. In this geometry the radiation phase depends on the horizontal electron position. We should take into account, that ODR is not an electron radiation, but the radiation of the target medium size of $\approx \gamma \alpha$. Each point of target radiate with its own phase. To account this feature we will start from the exact solution [3] of Maxwell equations for ODR from conductive semi-plane.

Let us $f(q)$ is any function one of $j_x(q)$, $j_z(q)$ and $\rho(q)$ from [3] (current and charge density), considered as function of $q$, where $q$ is a wave vector in z-direction. We should take into account, that the phase term in all expressions in [3], depending on the $z$ electron position $z_e$, is omitted. It is not necessary for [3], but we need it. So we will include it:

$$f(q) \cdot e^{iqz_e}$$

Here and lower we will use the space basis, shown in Fig.3.

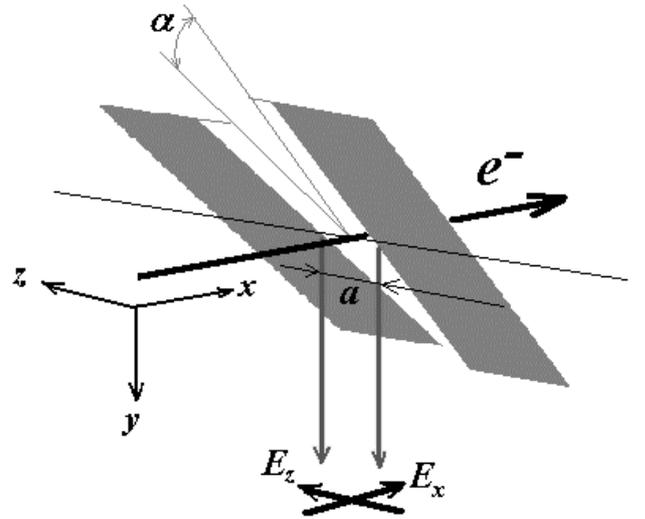

Fig.3. Radiation geometry for the artificial phase shift, depending on the transverse electron position

Let us turn into a dimensionless variable in z-direction:

$$\frac{1}{\sqrt{2\pi}} \int_{q=-\infty}^{\infty} f(q) \cdot e^{iqz_e} \cdot e^{-iqz} dq$$

Now we add the phase shift term, which phase shift depends on the z-variable and on the angle $\alpha$. After this we turn back into a wave vector space in z-direction

$$\frac{1}{\sqrt{2\pi}} \int_{z=-\infty}^{\infty} \frac{1}{\sqrt{2\pi}} \int_{q=-\infty}^{\infty} f(q) \cdot e^{iqz_e} \cdot e^{-iqz} dq \cdot$$
$$\cdot e^{i\omega(z-z_0)\alpha/\beta} \cdot e^{iq'z} dz =$$

$$= e^{iq'z_e} \cdot e^{i\omega(z_e-z_0)\alpha/\beta} \cdot f(q'+\omega\alpha)$$

The first term in the last equation is not important for next analysis and may be omitted. So we obtain the expression $f'(q)$ for the current and charge density in the semi-infinite plane, accounting the phase shift, evoked by the semi-plane turning.

$$f'(q) = e^{i\omega(z_e - z_0)\cdot\alpha/\beta} \cdot f(q + \omega\alpha) \qquad (1)$$

We can apply (1) for building expressions of ODR field from the distorted slit following the logics, used for example in [3] and [5]. Analyzing (1) we can see, that the ODR angular distribution will turn in the plane $x=0$ on the angle $\alpha$, because the argument $q$ in function $f$ is replaced by $q+\omega\alpha$. So the radiation from top and bottom semi-planes will diverge in this plane. This problem may be resolved by the using of a one-dimensional lens, which focus the full angular distribution in the plane x=0, because we are interesting the interference in the plane z=0. Lower we will assume this focusing is done and we will interest by angular distribution in $x$-$y$ plane only.

Following [5] and taking into account (1) we can write the expression for z-component of radiation field from suggested target in far zone for $\gamma \gg 1$:

$$E_z(\theta_z, \alpha, z, y) = E_z'(a/2 - z) \cdot e^{i\frac{2\pi}{\lambda}\alpha\cdot y} + E_z'^*(a/2 + z) \cdot e^{-i\frac{2\pi}{\lambda}\alpha\cdot y} \text{, where}$$

$$E'_z(h) = -i\frac{e}{4\pi^2} \cdot \frac{e^{-2\pi h/\lambda\left(\sqrt{\gamma^{-2}+\theta_x^2} + i\cdot\theta_z^2\right)}}{\sqrt{\gamma^{-2}+\theta_x^2} + i\cdot\theta_z^2}, \qquad (2)$$

$\theta_x$, $\theta_y$ are observation angles in units of $1/\gamma$ in respect to the direction of specular reflection, $a$ is a slit width, and $\{x, y\}$ is the electron position in respect to the slit center.

On Fig.4 the calculated using (2) angular distribution of ODR $z$-polarization component intensity as a function of the $y$ electron position for $\theta_x=0$, $z=0$, $\alpha=0.05$ and for the slit width $a=\gamma\lambda$ is shown.

We can see on Fig.4 that if the transversal electron position changes on some number of wavelength, the radiation angular distribution shifts at the angle $\approx 1/\gamma$. This effect may be used for the precise transversal beam size measurement.

Let us consider this possibility for the Gaussian transversal distribution of electrons in $y$ direction with the size parameter $\sigma$ and beam center $y_c$.

$$Y(\theta_z) \approx \int_y |E_z(\theta_z, \alpha, z, y)|^2 \frac{1}{\sqrt{2\pi\sigma^2}} \qquad (3)$$

Here we assume $y_c=0$ in the semi-planes cross point.

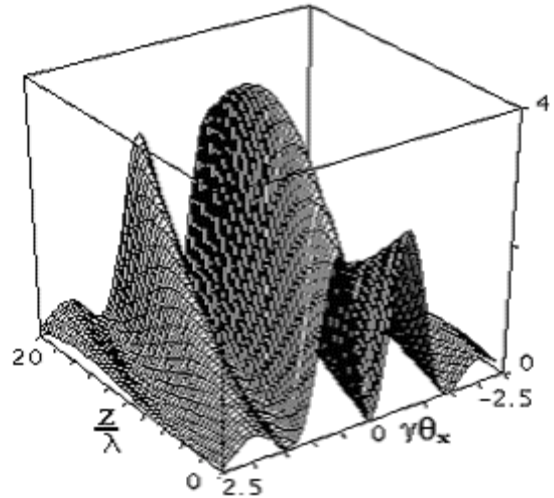

Fig.4 The calculated angular distribution of ODR $z$-polarization component intensity as a function of the $z$ electron position for $\theta_x=0$, $z=0$, $\alpha=0.05$ and for the slit width $a=\gamma\lambda$..

On Fig.5 and Fig.6 the angular distribution of ODR intensity $Y(\theta_z)$ as a function of the beam size $\sigma$ for two positions of the transversal beam center calculated using (3) is shown. We see on these figures, that the relation between minimum and maximum in angular distribution oscillations depends on $\sigma$, but the oscillation profile depends very on the electron beam position. It is difficult to separate a strong dependence on the beam size.

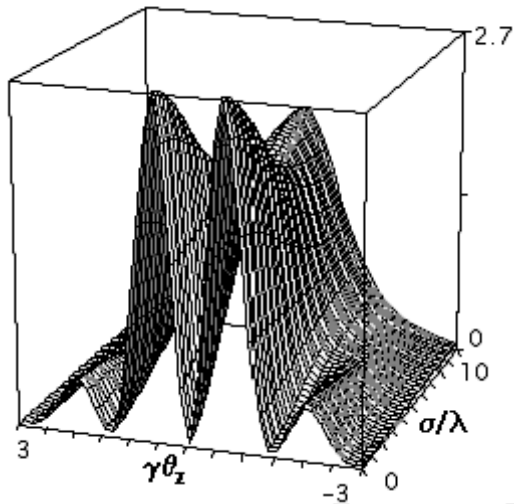

Fig.5. Angular distribution of ODR intensity $Y(\theta_z)$ as a function of the beam size $\sigma$ for $y_c=0$

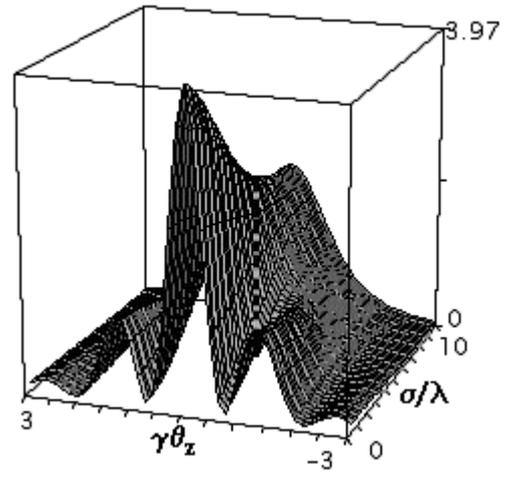

Fig.6. Angular distribution of ODR intensity $Y(\theta_z)$ as a function of the beam size $\sigma$ for $y_c=10\lambda$

To resolve this problem let us consider a new criteria $Y'(\theta_z)$.

$$Y'(\theta_z) = \sum_{k=-\infty}^{\infty} Y(\theta_z + kT),$$

where $T$ is the period of the angular distribution oscillation.
This functions being built for distributions shown on Figs. 5 and 6 is shown on Fig.7

The relation $\Delta$ between minimum and maximum of function $Y'(\theta_z)$ does not depend on the beam position $y_c$ and is convenient for the beam diagnostics. This relation is a strong function of the beam size (see Fig.8.). Using this dependence we can optimize the angle $\alpha$ to have a maximum sensitivity to the beam size.

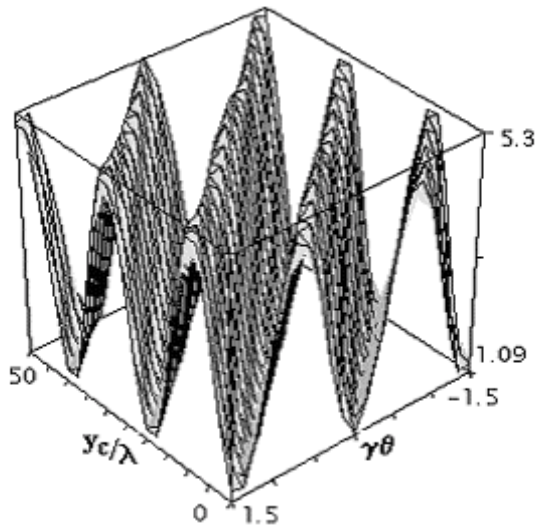

Fig.7. Criteria $Y'(\theta_z)$ as a function of beam position $y_c$ for a=$\gamma\lambda$, $\sigma$=3$\lambda$ and $\alpha$=0.05

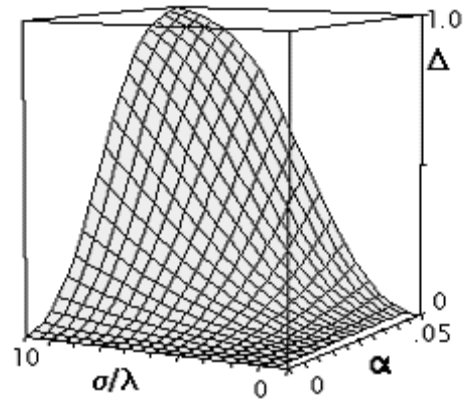

Fig.8. The relation $\Delta$ as a function of the beam size $\sigma$ and the distortion angle $\alpha$.

**Conclusion**
The method suggested allows us to chose the optimal sensitivity region so, that the response on the beam size may be comparable to the maximum ODR intensity. For the electron energy being equal to the ≈30GeV this value is at least 4 order larger, than the similar value in method, suggested in [2]. If we decrease the target slit width ($a<\gamma\lambda$) than the response on the beam size may increase at least up to transition radiation intensity. However in this case we should take into account the possible influence of the another transversal beam size in $z$ direction. In any case the sensitivity of this method is much more than the sensitivity in method suggested in [2].